\documentclass[twocolumn,showpacs,superscriptaddress,preprintnumbers,amsmath,amssymb,aps,prb]{revtex4-1}

\usepackage{graphicx}
\usepackage{dcolumn}
\usepackage{bm}
\usepackage{epsfig}
\usepackage{soul}
\usepackage{subfigure}
\usepackage{ulem}
\usepackage[usenames,dvipsnames]{color}

\begin{document}


\title{Magnon properties of random alloys }

\author{Fan Pan}
\affiliation{Department of Applied Physics, School of Engineering Sciences, KTH Royal Institute of Technology, Electrum 229, SE-16440 Kista, Sweden}
\affiliation{SeRC (Swedish e-Science Research Center), KTH Royal Institute of Technology, SE-10044 Stockholm, Sweden}
\author{Anna Delin}
\affiliation{Department of Applied Physics, School of Engineering Sciences, KTH Royal Institute of Technology, Electrum 229, SE-16440 Kista, Sweden}
\affiliation{SeRC (Swedish e-Science Research Center), KTH Royal Institute of Technology, SE-10044 Stockholm, Sweden}
\affiliation{Department of Physics and Astronomy, Materials Theory Division, Uppsala University, Box 516, SE-75120 Uppsala, Sweden}
\author{Anders Bergman}
\affiliation{Department of Physics and Astronomy, Materials Theory Division, Uppsala University, Box 516, SE-75120 Uppsala, Sweden}
\author{Lars Bergqvist}
\affiliation{Department of Applied Physics, School of Engineering Sciences, KTH Royal Institute of Technology, Electrum 229, SE-16440 Kista, Sweden}
\affiliation{SeRC (Swedish e-Science Research Center), KTH Royal Institute of Technology, SE-10044 Stockholm, Sweden}

\date{\today}

\begin{abstract}

We study magnon properties in terms of spin stiffness, Curie temperatures and magnon spectrum of Fe-Ni, Co-Ni and Fe-Co random alloys using a combination of electronic structure calculations and atomistic spin dynamics simulations. Influence of the disorder are studied in detail by use of large supercells with random atomic arrangement. It is found that disorder affects the magnon spectrum in vastly different ways depending on the system. Specifically, it is more pronounced in Fe-Ni alloys compared to Fe-Co alloys. In particular, the magnon spectrum at room temperature in Permalloy (Fe$_{20}$Ni$_{80}$) is found to be rather diffuse in a large energy interval while in Fe$_{75}$Co$_{25}$ it forms sharp branches. Fe-Co alloys are very interesting from a technological point of view due to the combination of large Curie temperatures and very low calculated Gilbert damping of $\sim$0.0007 at room temperature for Co concentrations around 20--30\%. 

\end{abstract}

\maketitle


\section{Introduction}

There has been a growing interest in disordered magnetic materials in the last few decades in the form of transition metal alloys and diluted magnetic semiconductors\cite{Moruzzi1990,Turek1994,Schroter1995,Abrikosov1996,James1999,Bergqvist2004,Sato2004,Ruban2005,Kudrnovsky2008,Sato2010,Polesya2010,Mankovsky2011}. A central motivation for many studies is the potential of these materials in spintronics and magnonics applications. Magnon excitations are commonly studied experimentally using inelastic neutron scattering suitable for bulk systems such as Co\cite{Sinclair1960} or spin polarized electron loss spectroscopy (SPEELS) for low dimensional magnets such as Co$_8$/Cu001\cite{Vollmer2003}. Theoretically, the simplest approach for calculating magnon spectrum for elements and compounds is through linear spin wave theory of the Heisenberg Hamiltonian. However, for accurate studies of alloys, both the treatment of disorder and thermal effects needs to be handled reliable. Magnons in disordered magnets, either random alloys or diluted, are more complicated than for ordered systems for a number of reasons. Due to broken translational symmetry, perfect magnon modes with infinite life time as in ordered magnets are absent but at certain conditions, one may still expect well defined magnon modes but with a finite lifetime due to disorder. The damping and formation of these modes are of great interest both theoretically and for applications. 

Previous studies of magnon properties of disordered magnets have been focused on diluted magnets and the effect of dilution on the magnon spectrum and spin stiffness\cite{Chakraborty2010,Chakraborty2015,Turek2016,Buczek2016}. The main findings from these studies are that the region with well defined magnon modes are decreasing with dilution and properties are strongly dimensionality dependent. Surprisingly, there are only very few published studies of magnons in random alloys with full concentration of magnetic elements\cite{Alben1977,Gennaro2018,Buczek2018}, such as Fe-Co alloys\cite{Masin2013,Schonecker2016}. The aim for the present study is to introduce a simple methodology for theoretical studies of magnons in disordered materials. We are using this methodology to investigate magnon and other finite temperature properties, i.e. spin stiffness, Cure temperatures and Gilbert damping for bulk transition metal alloys that hopefully will stimulate experiments in the new generation of neutron scattering facilities currently in construction.

The paper is organized as follows: In Section~\ref{sec:formalism} we introduce the methodology and give the details of the calculations, in Section~\ref{sec:Results} we present our findings and in Section~\ref{sec:summary} we give a summary and provide an outlook.

\section{Formalism} \label{sec:formalism}

\subsection{Spin excitations in solids}   

A  magnetic solid at finite temperature displays two different kinds of magnetic excitations, namely spin wave excitations (magnons) and electron-hole pair excitations (Stoner). The magnon excitations are responsible for transversal fluctuations while Stoner excitations cause longitudinal changes of the moments. At low temperatures and in particular for bulk materials, as in this study, the magnon excitations dominate and as a first approximation the Stoner excitations can be neglected.  
However, it is worth noting that they may play an important role at high temperatures and also for certain materials with induced magnetic moments. Longitudinal fluctuations can however be modelled in a more advanced treatment\cite{Pan2017}.
 
The low energy spin excitation in a form of a magnon is characterized by the wave vector $\mathbf{q}$ within the Brillouin zone and for a cubic, ordered, material the magnon energy $E(\mathbf{q}) = \hbar \omega(\mathbf{q}) \approx D q^2$, where $D$ is the spin wave stiffness\cite{Pajda2001} 

\begin{equation}
D= \frac{2}{3M} \sum_j J_{0j} \mathbf{R}_{oj}^2,
\end{equation}
and $M$ is the magnetization, $J_{ij}$ is the exchange interactions between magnetic moments $\mathbf{m}$ at sites $i$ and $j$ connected with position vector $\mathbf{R}$. In the case of disorder, the spin wave stiffness $D$ is obtained in by averaging over all $N$ atoms in the system as 
\begin{equation}
D= \frac{2}{3M} \frac{1}{N} \sum_n \sum_j J_{nj} \mathbf{R}_{nj}^2,
\end{equation}  
  
\subsection{Atomistic spin dynamics}

The dynamics of a magnetic material at finite temperature and thus the magnetic excitations, is conveniently modelled through atomistic spin dynamics (ASD) simulations\cite{ASDbook2017}.  Within ASD, the temporal evolution of the atomic moments $\mathbf{m}$ at finite temperature is governed by Langevin dynamics, through coupled stochastic differential equations, the Landau-Lifshitz-Gilbert (LLG) equations, here written in the Landau-Lifshitz form, 

\begin{eqnarray}
\frac{d \mathbf{m}_i}{d t} & = & - \frac{\gamma}{(1+\alpha ^2)}  \mathbf{m}_i \times \left[\mathbf{B}_i + \mathbf{b}_i(t)\right]  \\ \nonumber   & - & \gamma\frac{\alpha}{m (1+\alpha ^2)} \mathbf{m}_i \times \left\{ \mathbf{m}_i \times\left[\mathbf{B}_i + \mathbf{b}_i(t)\right] \right\},
\label{eqn:LLG}
\end{eqnarray}
\noindent where $\gamma$ is the electron gyromagnetic ratio and $\alpha$ is the Gilbert damping parameter. The latter can either be taken from experiments using ferromagnetic resonance (FMR) or calculated from first-principles. The effective interaction field $\mathbf{B}_i$ experienced by each atomic moment $i$ is given by 

\begin{equation}
\mathbf{B}_i = - \frac{\partial\mathcal{H}}{\partial\mathbf{m}_i}.
\end{equation}
where $\mathcal{H}$ is the spin Hamiltonian governing the interactions between the magnetic moments. We are employing the semi-classical Heisenberg model, \\ $\mathcal{H}=-\sum_{ij} J_{ij} \mathbf{m}_i \cdot \mathbf{m}_j $, where the exchange interactions are parametrized from first-principles calculations. The effective interaction field is complemented with a stochastic field $\mathbf{b}_i$ that is modeled with uncorrelated white noise with a temperature dependent variance\cite{ASDbook2017}.

\subsection{Magnon dispersion} \label{sec:MD}

We are employing two different complementary methods for calculating the magnon spectrum, 1) the adiabatic magnon spectrum (AMS) valid for the ground state and 2) from ASD simulations through the dynamical structure factor at finite temperatures and damping.

\subsubsection{Adiabatic magnon spectrum}

The adiabatic magnon spectrum is directly connected to the real-space exchange interactions $J_{ij}$ through Fourier transformation\cite{Anderson1963,Halilov1998}. Let $J^{\alpha\beta}(\mathbf{q})$ denote the Fourier transform of the exchange interaction between chemical type $\alpha$ and $\beta$ with a wave-vector $\mathbf{q}$ lying in the Brillouin zone (BZ). $J^{\alpha\beta}(\mathbf{q})$ is calculated as

\begin{equation}
J^{\alpha\beta}(\mathbf{q})=\sum_{j\ne0} J_{0j}^{\alpha\beta} \mathrm{exp}(i \mathbf{q} \cdot \mathbf{R}_{0j}).
\end{equation}

In the spirit of virtual crystal approximation (VCA), it is tempting to perform a chemical average of the Fourier transformed exchange interactions, i.e.

\begin{equation}
 \tilde{J}(\mathbf{q})=J^{11}(\mathbf{q})x_1^2+J^{12}(\mathbf{q})x_1x_2+J^{21}(\mathbf{q})x_1x_2+J^{22}(\mathbf{q})x_2^2
 \end{equation} 
in the case of binary alloy and where $x_1$ and $x_2$ are the concentration of each chemical type. In such a case, the "effective" magnon energy ($\hbar$=1) for each wavevector $\mathbf{q}$ can then be adapted to the expression valid for one atom/cell of ordered systems\cite{Anderson1963,Halilov1998}

\begin{equation} \label{eq:AMS}
\tilde{\omega}(\mathbf{q})=\frac{4}{\tilde{M}} \left( \tilde{J}(\mathbf{0})-\tilde{J}(\mathbf{q}) \right), 
\end{equation}
where $\tilde{M}$ is the saturation magnetization. However, this treatment of the disorder is over-simplified and does not reproduce experimentally observed excitations. Analogous to multi-sublattice ordered systems, where $N$ magnon branches appear in the spectrum ($N$ is the number of sublattices), chemically disordered systems containing $K$ chemical components will exhibit $K$ magnon branches. More specifically, in the case of a binary alloy ($K$=2), the magnon energies at each wave-vector $\mathbf{q}$ will be given by the eigenvalues of the following $2 \times 2$ dynamical matrix

\begin{equation} \label{eq:AMS2}
\omega(\mathbf{q})=  
4 \mathrm{Eig} 
\left( \begin{array}{cc} 
\frac{ (J^{11}(\mathbf{0})-J^{11}(\mathbf{q}))x_1+J^{12}(\mathbf{0})x_2}{M_1} &
 -\frac{J^{12}(\mathbf{q})x_2}{M_1} \\  
 -\frac{J^{21}(\mathbf{q})x_1}{M_2} &  
 \frac{ (J^{22}(\mathbf{0})-J^{22}(\mathbf{q}))x_2+J^{21}(\mathbf{0})x_1}{M_2} \end{array} \right ).
\end{equation}

\subsubsection{Dynamical structure factor} \label{sec:SQW}

 The magnon dispersion at finite temperatures are directly accessible in ASD through the dynamical structure factor $S(\mathbf{q},\omega)$ \cite{Bergman2010,Bergqvist2013,Etz2015}. 
 The key ingredient is the measurement of the time and space correlation function

\begin{equation} 
\label{c_k}
C^{\mu \nu}(\mathbf{r},t)=\frac{1}{N}\sum_{\substack{i,j ~ \text{where} \\\ ~\mathbf{r}_i-\mathbf{r}_j=\mathbf{r}}} \langle m_i^\mu(t) m_j^\nu(0) \rangle -  \langle m_i^\mu(t) \rangle \langle m_j^\nu(0) \rangle.
\end{equation}
The correlation function defined in Eqn.~(\ref{c_k}) describes how the magnetic order evolves both in space ($\mu , \nu$ denotes carteisian components) and over time. The perhaps most valuable application of $C(\mathbf{r},t)$ is obtained by a Fourier transform over space and time to give the dynamical structure factor 
\begin{equation} 
\label{sqw}
S^{\mu \nu}(\mathbf{q},\omega)=\frac{1}{\sqrt{2\pi}N}\sum_{\mathbf{r}} e^{i\mathbf{q}\cdot\mathbf{r}} \int_{-\infty}^\infty e^{i\omega t}C^{\mu \nu}(\mathbf{r},t) \,dt.
\end{equation}

The magnon energies are determined by the peak values of $S(\mathbf{q},\omega)$ at wavevector $\mathbf{q}$. In contrast to the adiabatic treatment, temperature effects from the Gilbert damping processes are included that give rise to intensity variation of the available energies. 
In the present study, we have not included longitudinal fluctuations of the magnetic moment. Such fluctuations give rise to Stoner excitations and an additional damping mechanism for magnons – so-called Landau damping.

\subsection{Details of calculation} \label{sec:calcdetails}

All first-principles calculations in this study was performed using a multiple-scattering (Korringa-Kohn-Rostoker, KKR) implementation of the density functional theory (DFT) as implemented in the SPR-KKR software \cite{Ebert2011a,SPRKKR}. The generalized gradient approximation (GGA) using the Perdew-Burke-Enzerhof (PBE) parametrization was used as exchange-correlation for the volume relaxation while all other calculations employed the local spin density approximation (LDA).
 The calculations are fully relativistic employing the atomic sphere approximation with a basis set consisting of $spdf$-orbitals. The coherent potential approximation (CPA) was employed for treating the disorder. In order to study the magnetic excitations and finite temperature properties, the total energies from the electronic structure calculations are mapped onto an effective Heisenberg Hamiltonian generalized to random alloys.

The magnetic exchange interactions were obtained from the magnetic force theorem using the formalism of Lichtenstein, Katsnelson, Antropov and Gubanov (LKAG)\cite{LKAG1984,LKAG1987}. Gilbert damping was calculated using the linear response formalism of the torque-torque correlation method as described in Ref.[\onlinecite{Mankovsky2013}]. The alloy-analogy model within CPA \cite{Ebert2015} was employed for the finite temperature damping where both atomic displacements and spin fluctuations from Monte Carlo data were included. The atomistic simulations, either the Monte Carlo or atomistic spin dynamics, were performed using the UppASD software \cite{ASDbook2017,UppASDcode}. Here the disorder is instead treated by using a large supercell in which each site is chemically randomly occupied according to the concentration. We are using large supercells consisting of between 110592 atoms (for the calculation of the spin stiffness and AMS) and 512000 atoms (for the calculation of the dynamical structure factor), such that most of the local environment configurations from a central atom exist within the supercell.
The spin stiffness was calculated for each individual atom in the supercell and the final result was obtained by performing an average over all atoms.

\section{Results} \label{sec:Results}

\subsection{Electronic band structure}

\begin{figure}[htp] 
\includegraphics[width=9cm]{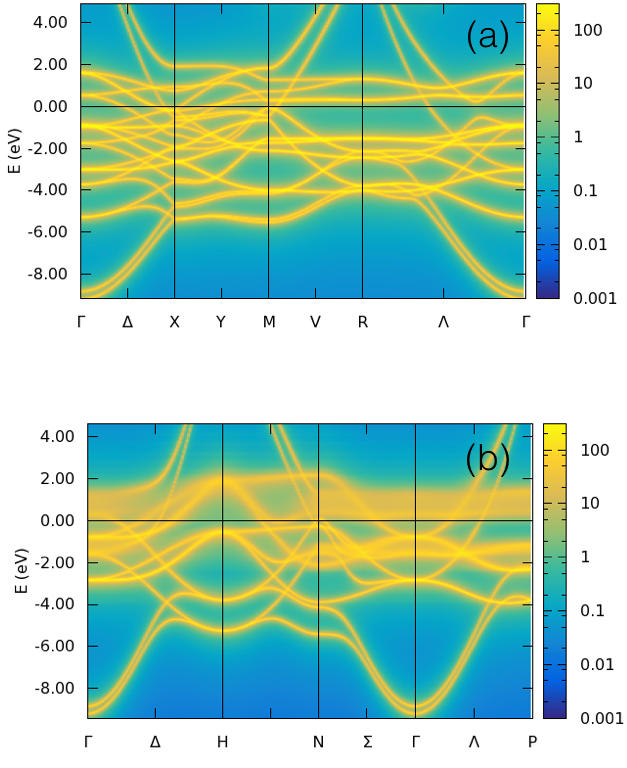}
\caption{Electronic band structure in terms of the Bloch spectral function of a) ordered Fe-Co compound in the B2 structure and b) Fe$_{50}$Co$_{50}$ random alloy in the bcc structure.}
\label{fig:BSF}
\end{figure}

Before describing in details how the magnetic properties are affected by chemical disorder, we first look into the electronic band structure. For ordered elements and compounds, the electron bands are well defined with no associated broadening as function of energy and wave vector within the LDA/PBE treatment. This corresponds to electrons having infinite lifetime. A typical electron band structure of an ordered Fe-Co compound is displayed in Fig.~\ref{fig:BSF}a). However, if the system has chemical disorder (or if the finite lifetime of the quasi-particles are taken into account) the bands become "fuzzy" and obtain a finite broadening, with a line width inversely proportional to the lifetime. The broadening is however not uniform over the considered energy range. For random alloys, the electron band structure is conveniently obtained through the Bloch spectral function within CPA, as demonstrated in Fig.~\ref{fig:BSF}b) where the electron band structure of Fe$_{50}$Co$_{50}$ random alloy in the body centered cubic (bcc) lattice  is displayed. For this particular system and concentration, the disorder is most visible around the Fermi level . This affects many properties such as the Gilbert damping (see Section~\ref{sec:damping}).

\subsection{Spin stiffness} \label{sec:spin stiffness}

\begin{figure}[htb]
\includegraphics[width=9cm]{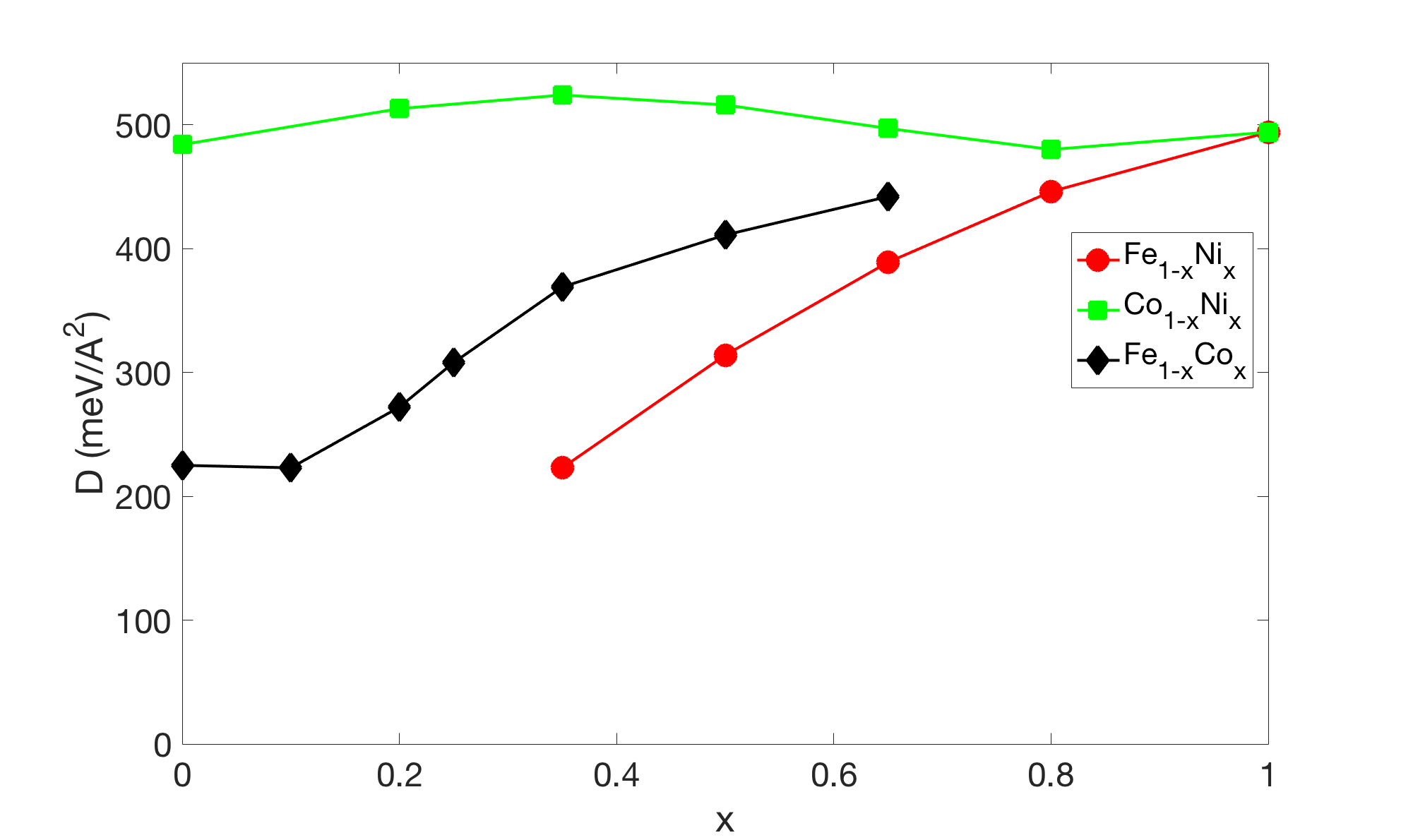}
\caption{Calculated spin stiffness $D$ in (meV\AA$^2$) for the random alloys Fe$_{1-x}$Ni$_x$, Co$_{1-x}$Ni$_x$ and Fe$_{1-x}$Co$_x$.}
\label{fig:stiffness}
\end{figure}

The calculated values of the spin stiffness $D$ are shown in Fig.~\ref{fig:stiffness} for Fe-Ni, Co-Ni and Fe-Co random alloys. For the Fe-Ni alloys, $D$ is monotonously increasing with the Ni concentration.  This has a rather simple explanation. Pure Fe in the face-centered cubic (fcc) lattice at the here considered volumes possess rather complicated non-collinear magnetic structures\cite{Abrikosov1999} which translates into a vanishing spin stiffness. Overall, the magnetic properties in Fe-Ni alloys are rather sensitive to volume changes. Even at Invar concentration (Fe$_{65}$Ni$_{35}$) it is possible to kill the ferromagnetic order by applying pressure, and in this way obtain a vanishing spin stiffness.

The Co-Ni alloys behave differently. Here the values of the spin stiffness are rather constant throughout the whole concentration range and therefore the magnon properties are not expected to change much. This is perhaps not so unexpected since both elemental Co and Ni are stable in the fcc lattice.

The spin stiffness of the Fe-Co alloys in the bcc lattice shows a more interesting behaviour. At low concentrations of Co ($x< 0.2$), the spin stiffness is similar as for elemental Fe while it increases for higher concentrations of Co. At the phase boundary around $x=0.7$, the spin stiffness is approximately twice as large as that of Fe. This suggests that the are ample possibilities for tuning the magnetic properties in this system.

\subsection{Curie temperatures}

\begin{figure}[htb]
\includegraphics[width=9cm]{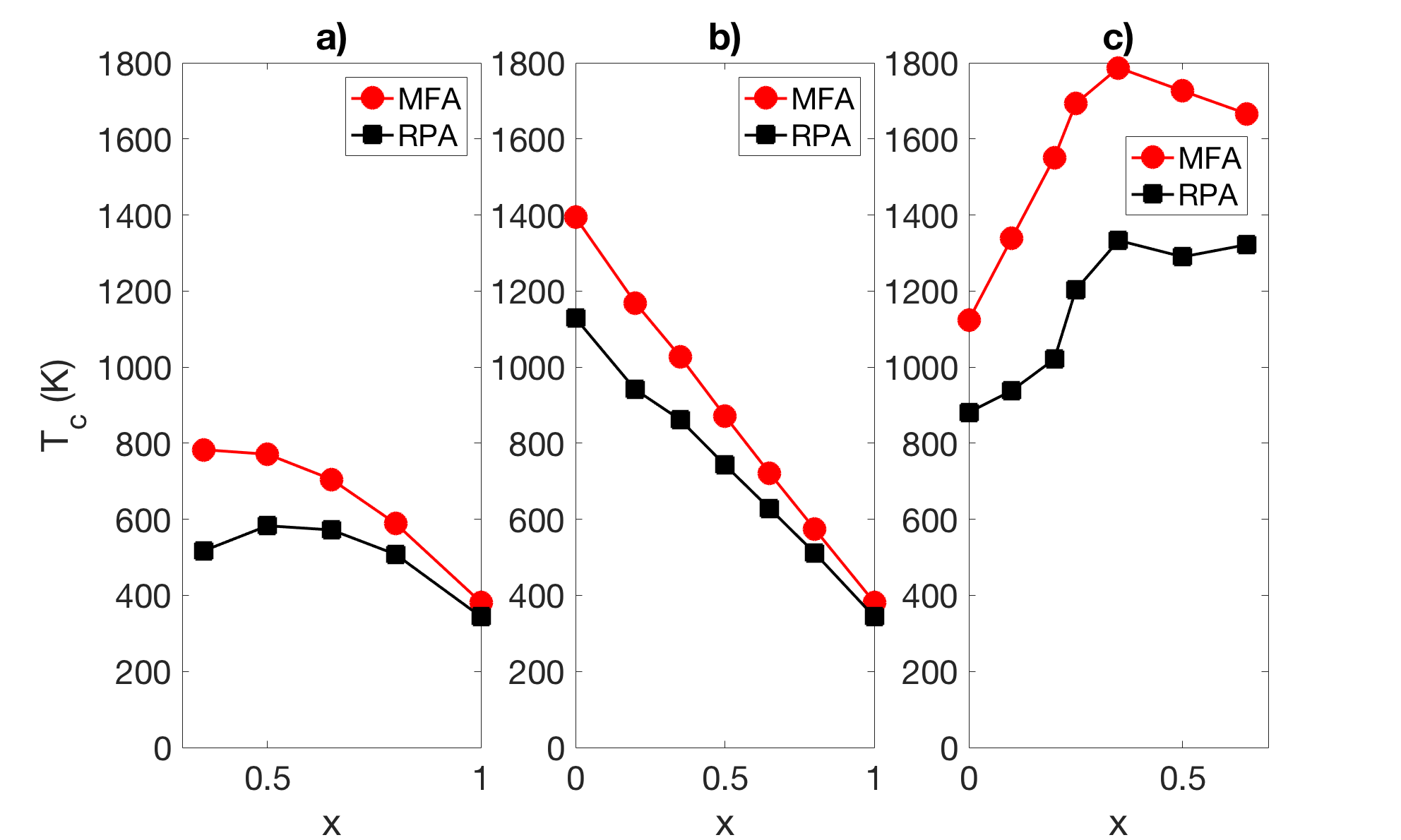}
\caption{Calculated Curie temperatures for the random alloys a) Fe$_{1-x}$N$i_x$ ), b) Co$_{1-x}$Ni$_x$ and c) Fe$_{1-x}$Co$_x$. MFA denotes values from mean field approximation and RPA from random phase approximation.}
\label{fig:tc}
\end{figure}

Our computed Curie temperatures are shown in Fig.~\ref{fig:tc}. Two different approaches have been used, the mean field approximation (MFA) and the random phase approximation (RPA). In principle, MFA corresponds to the arithmetric average of the exchange interactions and RPA to the harmonic average. It can be shown\cite{Majlis2001,Sinai1982} that for ferromagnetic interactions $T_c^{MFA} > T_c > T_c^{RPA}$, where $T_c$ is the "true" value (which can be obtained from Monte Carlo). The two different methods (MFA and RPA) then set the upper and lower bounds of $T_c$. Of the three considered alloy systems, the Fe-Ni system displays the lowest values of $T_c$ while Fe-Co the highest with values peaking around 1500 K for Co concentrations around 0.5. 

\subsection{Gilbert damping} \label{sec:damping}

\begin{figure}[htp]
\includegraphics[width=9cm]{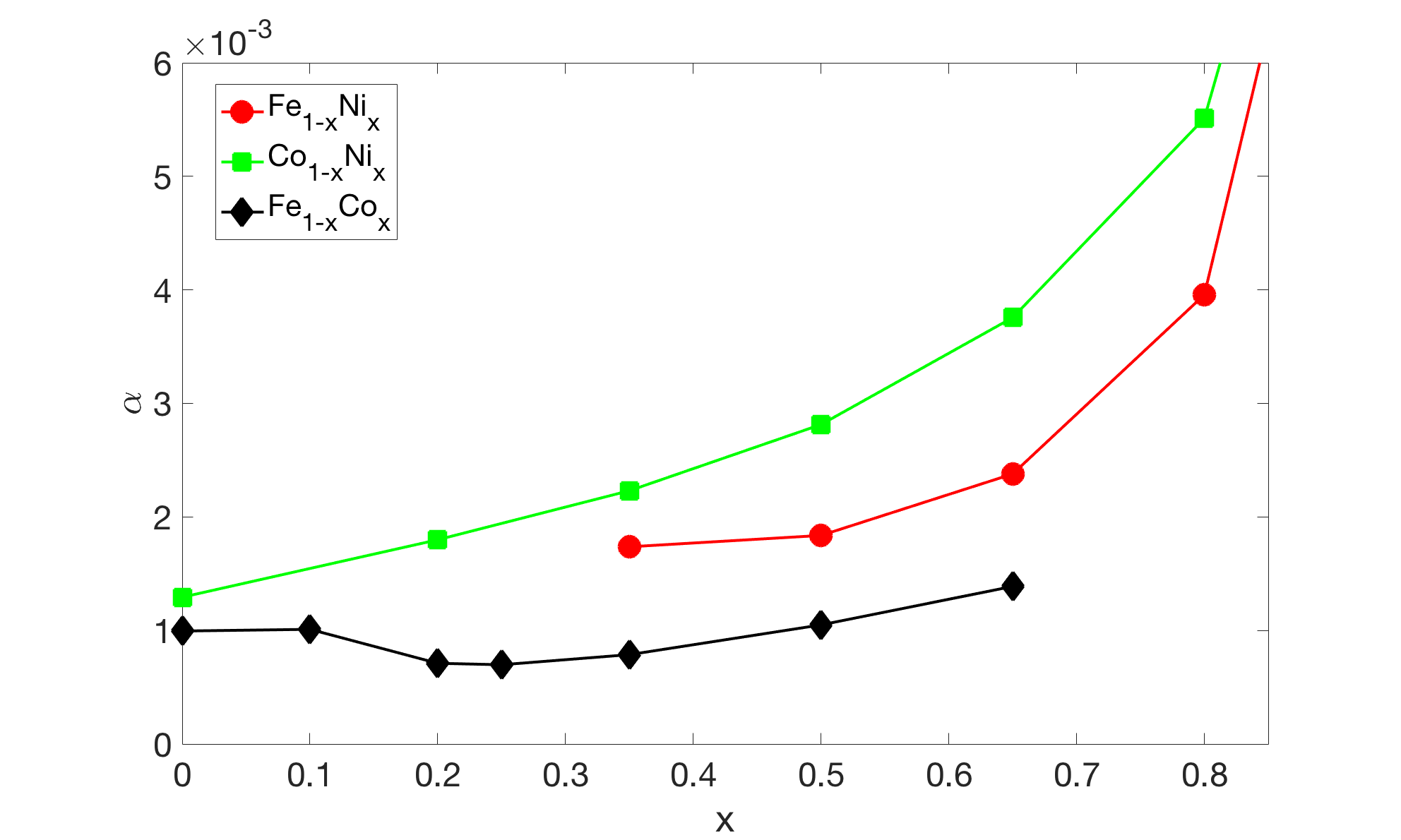}
\caption{Calculated Gilbert damping at $T=$ 300 K for the random alloys Fe$_{1-x}$N$i_x$ (red), Co$_{1-x}$Ni$_x$ (green) and Fe$_{1-x}$Co$_x$ (black).}
\label{fig:damping}
\end{figure}

Gilbert damping in magnetic materials determines the rate of dissipative energy processes with the surroundings. Very often a low damping is wanted in order to minimize energy losses but equally important is the ability to tune the damping. This can be achieved by, e.g., impurity doping\cite{Pan2016} or by varying the alloy composition. The latter is pursued here. For both fcc alloy systems considered in this study, i.e. Fe-Ni and Co-Ni, the Gilbert damping increases with Ni concentration. The Gilbert damping for Fe-Ni is consistently lower than the one seen in Co-Ni.  For elemental Ni (off scale), we obtain $\alpha=0.013$, which is in the same range as reported previously \cite{Mankovsky2013,Gilmore2007}.  Worth noting is that the damping is one order of magnitude smaller in elemental Co and Fe. What is perhaps most remarkable however is the very low damping found in certain Fe-Co alloys, in which it is even lower than for elemental Fe. This behaviour is due to variation of the density of states and was explained in detail in previous studies\cite{Mankovsky2013,Schoen2016}. The experimental values reported in Ref.[\onlinecite{Schoen2016}] are in good agreement with our calculated values presented here.

\subsection{Magnon properties}

\begin{figure}[htp]
\includegraphics[width=8cm]{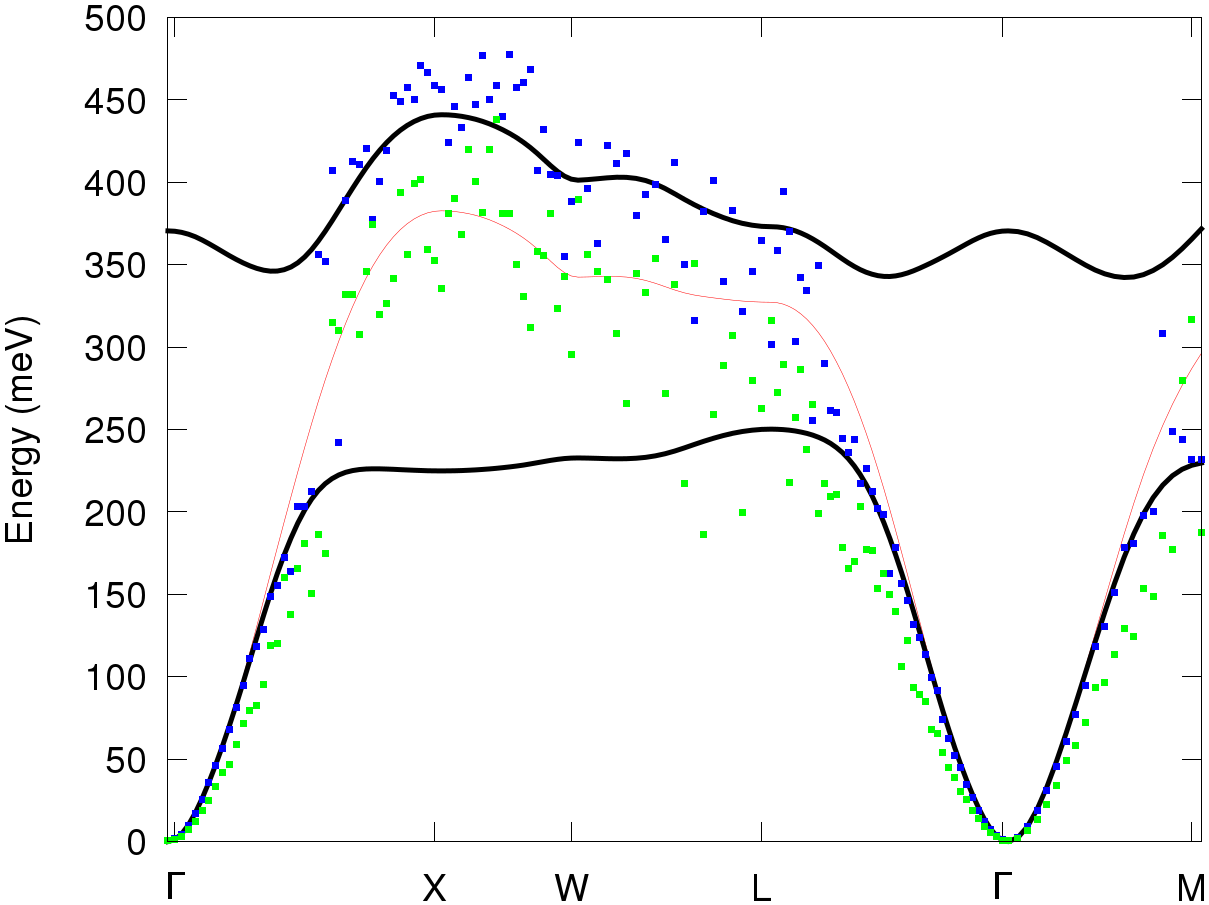}
\caption{Magnon spectrum of permalloy (Fe$_{20}$Ni$_{80}$). The thin red line denotes effective adiabatic spectrum, Eq.(\ref{eq:AMS}), and thick black lines full adiabatic treatment, Eq.(\ref{eq:AMS2}). Blue (green) points denote peak position at each wavevector of the dynamical structure factor from atomistic spin dynamics calculations at
$T =$ 10 K  ($T =$ 300 K) using the calculated Gilbert damping.}
\label{fig:AMSPy}
\end{figure}

Overall, we find that the main features of the magnon spectra are quite similar in all systems we have considered here. We therefore choose in this section to present results only for two systems of particular technological interest: i) permalloy (Fe$_{20}$Ni$_{80}$) in the fcc lattice and ii) Fe$_{75}$Co$_{25}$  in the bcc lattice chosen due to its large magnetic moment and low damping. 

\begin{figure}[htb]
\includegraphics[width=9cm]{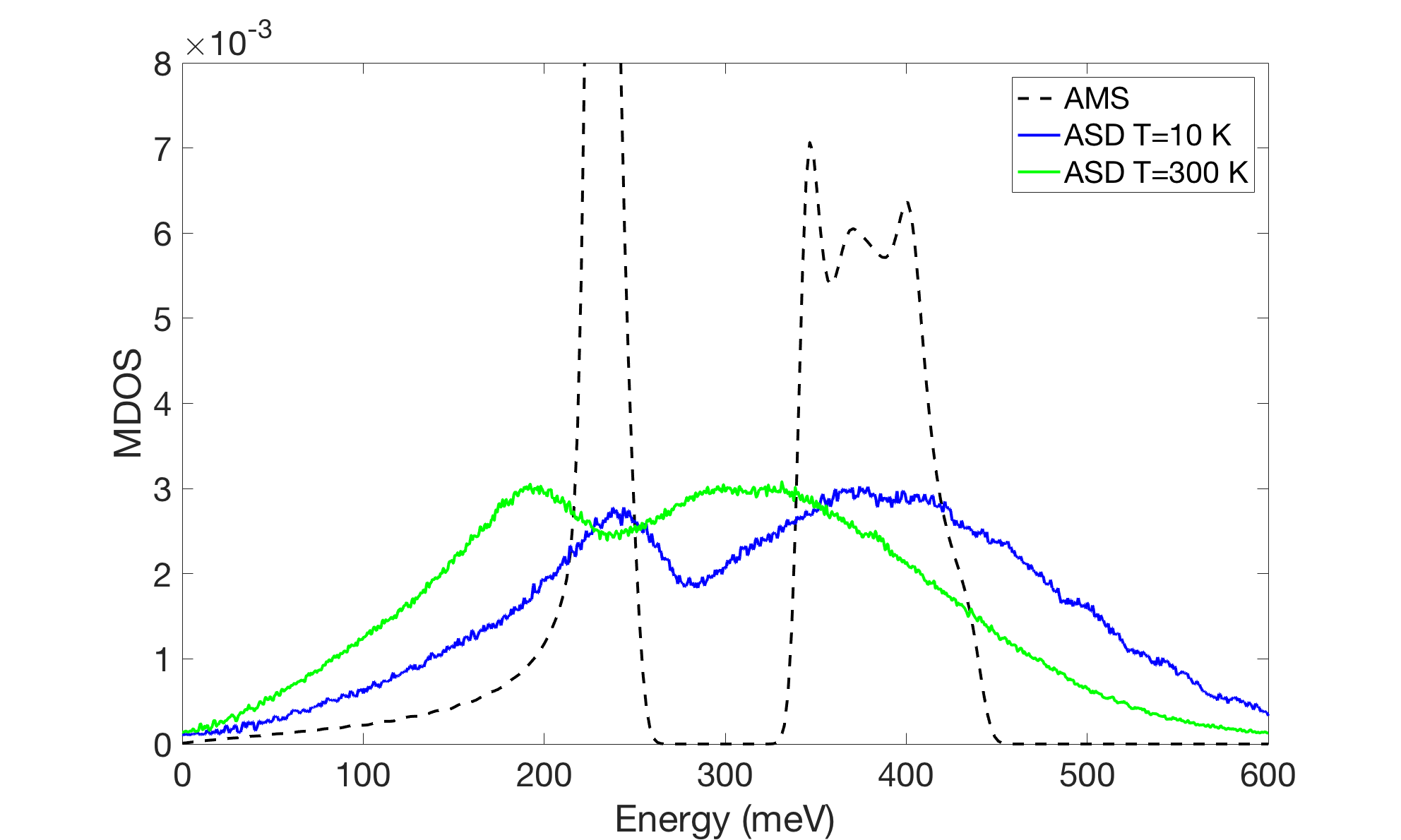}
\caption{Magnon density of states of permalloy (Fe$_{20}$Ni$_{80}$) from AMS and atomistic spin dynamics simulations at $T =$ 10 K and $T =$ 300 K. }
\label{fig:PyMDOS}
\end{figure}

In Fig.~\ref{fig:AMSPy}, the calculated magnon spectrum of Py is displayed using a variety of different tools as described in Section.~\ref{sec:MD}. Both the thin red line and the bold black lines are from adiabatic calculations, Eqs.~(\ref{eq:AMS}) and (\ref{eq:AMS2}), respectively. Between the two, the spectrum in black lines is expected to hold which is clear from comparison with dynamical structure factor from atomistic spin dynamics calculations as indicated in squares for two different temperatures, namely $T =$ 10 K and $T =$ 300 K.  It is important to remember that AMS only reflects the exchange interactions and the chemical disorder of the system. Temperature effects in the form of transversal fluctuations and damping are however included in the ASD simulations where the calculated Gilbert damping at $T=$ 300 K was employed. The curvature around the $\Gamma$ point is the spin stiffness and by inspection it is clear that the spectrum softens drastically at room temperature compared to the low temperature data. This temperature dependence of the stiffness was also analyzed in a recent study\cite{Gennaro2018}. At higher energies, due to a combination of thermal fluctuations, disorder and damping processes the spectrum broadens which is much clearly shown in Fig.~\ref{fig:PyMDOS} where the magnon density of states (MDOS) is displayed.  

\begin{figure}[htp]
\includegraphics[width=8cm]{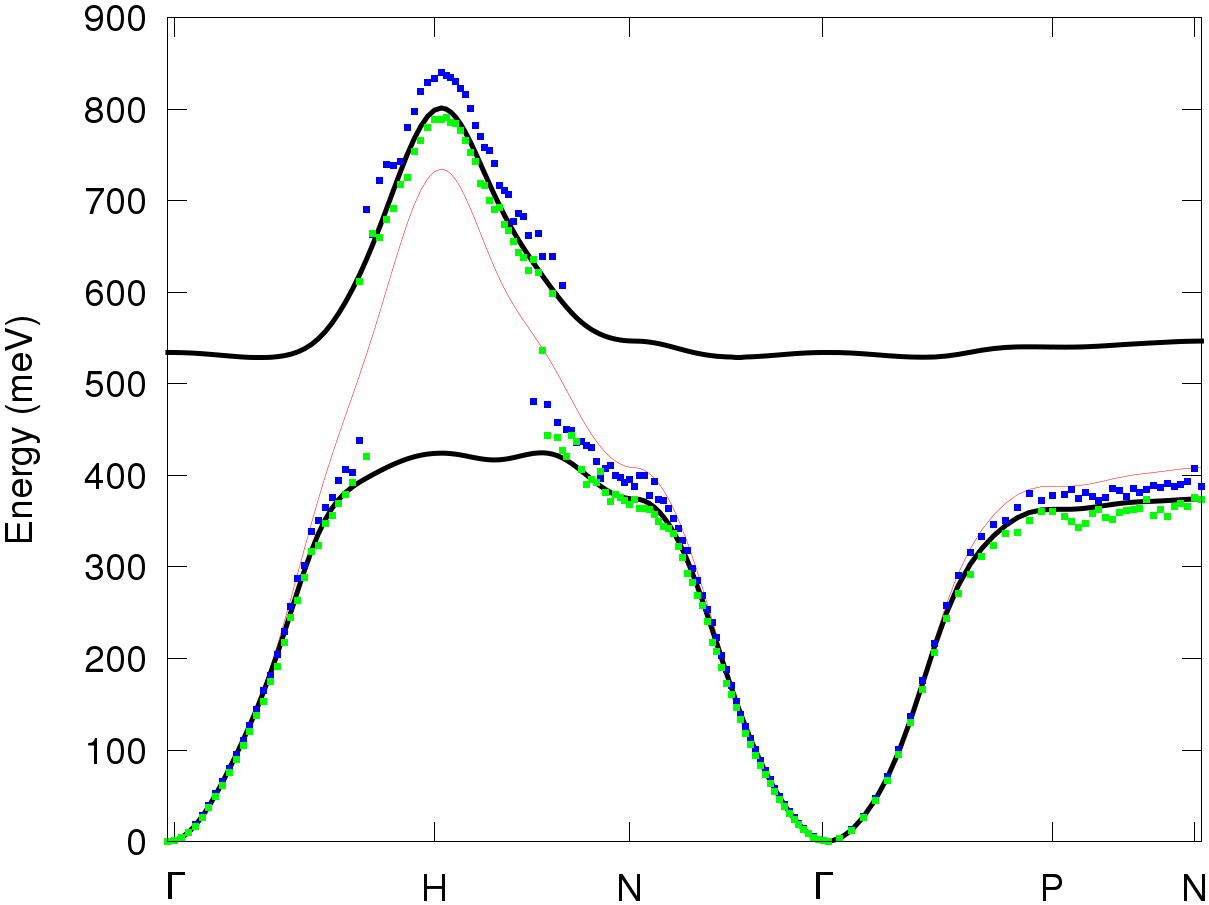}
\caption{Magnon spectrum of Fe$_{75}$Co$_{25}$. The thin red line denotes effective adiabatic spectrum, Eq.~(\ref{eq:AMS}), and thick black lines full adiabatic treatment, Eq.~(\ref{eq:AMS2}). Blue (red) points denote peak position at each wavevector of the dynamical structure factor from atomistic spin dynamics calculations at $T =$ 0 K ($T =$ 300 K) using the calculated Gilbert damping. }
\label{fig:AMSFeCo}
\end{figure}

The MDOS obtained from AMS has two distinct peaks which we can denote "acoustic" and "optical" branch in analogy to phonons. The two branches are separated by a small gap. However, even at low temperature ($T =$ 10 K), the MDOS as obtained by ASD simulations is broadened enough such that the two branches overlap. The peak positions are however almost identical. Although one need to keep in mind that AMS is using a simplified treatment of disorder, namely VCA, while ASD simulations are treating the disorder much more accurately by using a large random supercell. Increasing the temperature to 300 K softens the spectrum almost uniformly. This finding has been used to describe the low temperature dependence of MDOS with a quasiharmonic approximation in Refs.~[\onlinecite{Woo2015,Bergqvist2018}].

The calculated magnon spectrum for Fe$_{75}$Co$_{25}$, displayed in Fig.~\ref{fig:AMSFeCo}, is quite different from the one of Py. First of all, given its much higher Curie temperature the difference of the spectrum between $T =$ 10 K and $T =$ 300 K is minimal. Secondly, since Fe and Co atoms are rather similar chemically, both in terms of magnetic moments,  2.5$\mu_B$ and 1.8$\mu_B$, respectively and the exchange interactions are of similar magnitude, the spectrum has much less disorder broadening. 

\subsection{Magnon lifetimes, ordered versus disordered}

\begin{figure}[hbp]
\includegraphics[width=8cm]{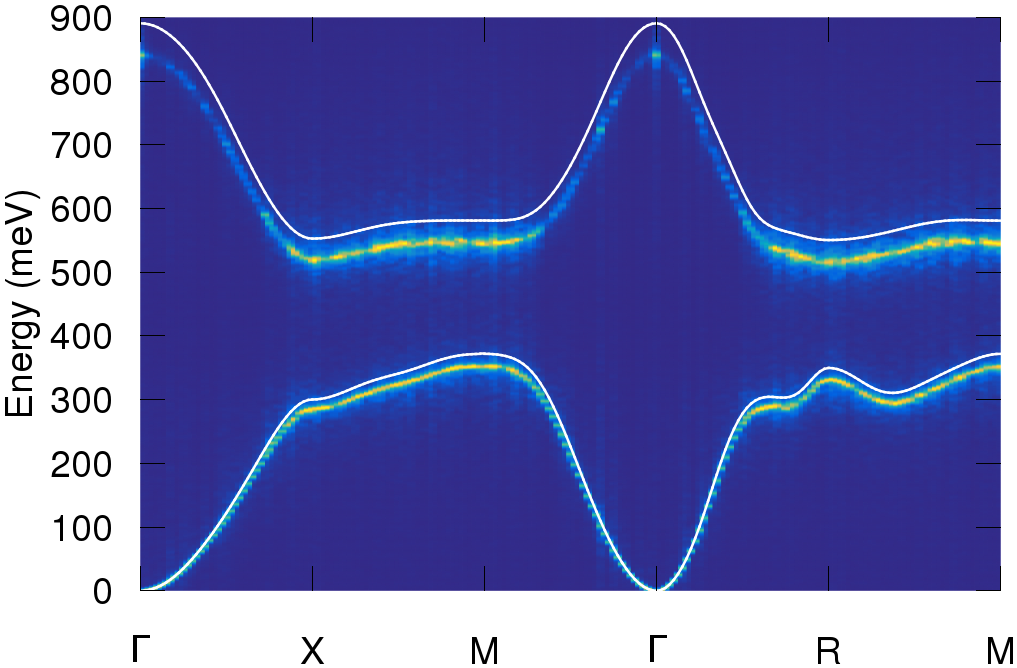}
\includegraphics[width=8cm]{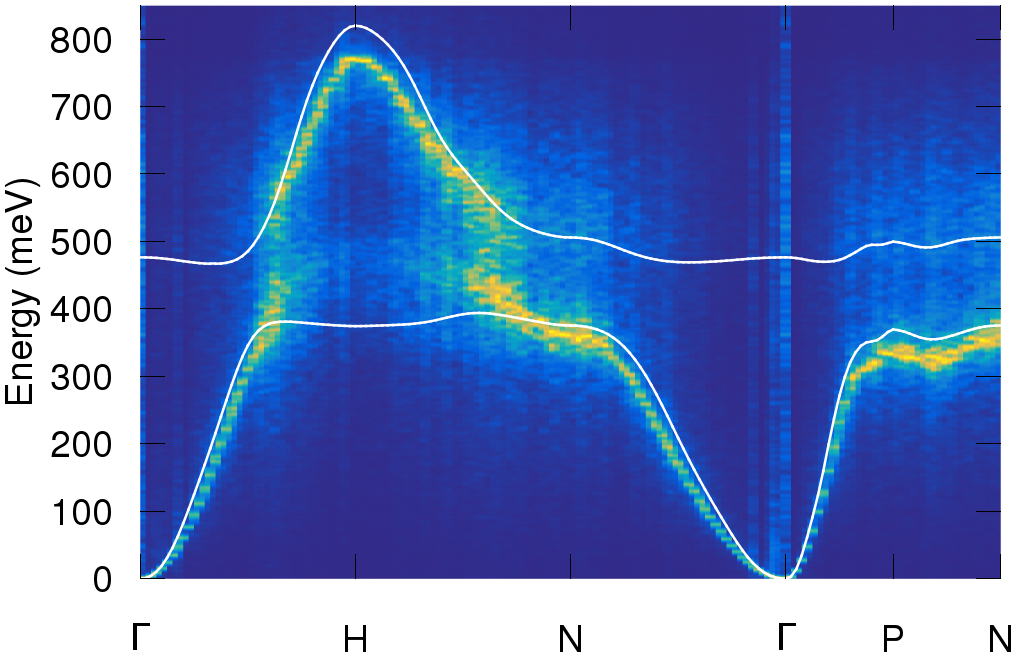}
\caption{Magnon spectrum from atomistic spin dynamics simulations at $T =$ 300 K of Fe$_{50}$Co$_{50}$ in the ordered B2 structure (top) and as a random alloy (bottom). The white line is the corresponding spectrum as obtained from AMS.  }
\label{fig:FeCoSQW}
\end{figure}

To further quantify the effects of disorder on magnon properties, in this section we compare ordered system with disordered system having the same composition. More specifically, we compare Fe$_{50}$Co$_{50}$ which exists both in ordered structure, B2, or as a disordered random alloy in bcc structure. Both the magnetic moments ($M_s \approx 2.2 \mu_B$) and Curie temperatures ($\approx $ 1400 K) are rather similar between the two structures. The calculated Gilbert damping at room temperature is however lower for the ordered B2 structure, 0.0007 vs 0.0011 for the random alloy. It is worth noting that the damping for the B2 is remarkable low for a metallic compound. In Fig.~\ref{fig:FeCoSQW}, the magnon spectrum from ASD simulations at $T =$ 300 K is shown for the both compounds, together with the AMS spectrum as reference. Due to the lack of disorder in the B2 structure, the magnon states are very well defined throughout the whole Brillouin zone. It is immediately clear that the disorder of the random alloy broadens the magnon states affecting the magnon lifetimes, similar as found for the electron bands in Fig.~\ref{fig:BSF}. However, even for the random alloy there are relatively well defined magnon states throughout the Brillouin zone, in contrast to the Fe-Ni alloys where the magnon states away from the $\Gamma$-point are very diffuse.

\begin{figure}[htp]
\includegraphics[width=9cm]{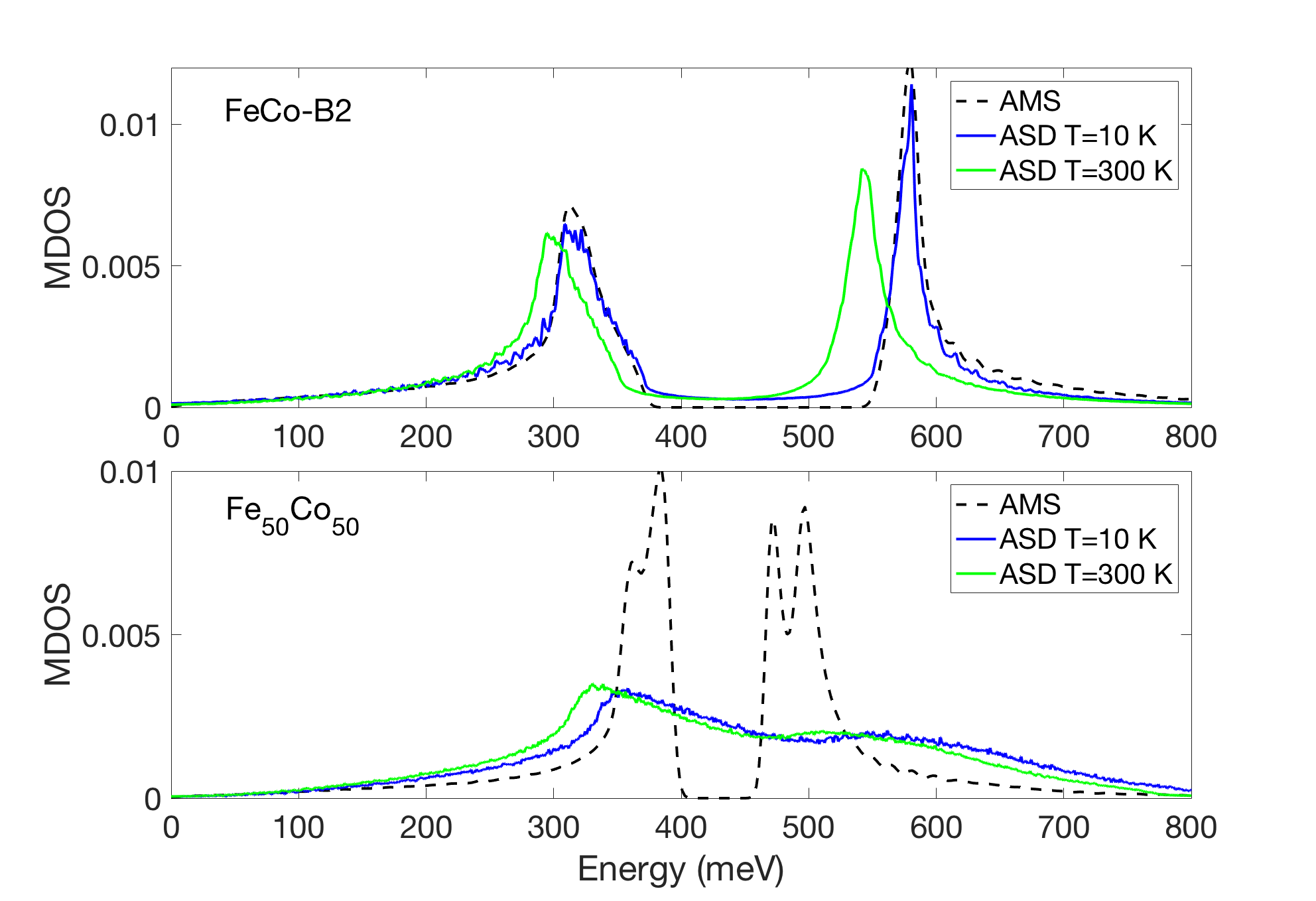}
\caption{Magnon density of states of Fe$_{50}$Co$_{50}$ in ordered B2 structure (top) and random alloy (bottom) from AMS and atomistic spin dynamics simulations at $T =$ 10 K and $T =$ 300 K. }
\label{fig:FeCoMDOS}
\end{figure}

In Fig.~\ref{fig:FeCoMDOS}, the magnon DOS is displayed for the two compounds. As also clear from the spectrum, in the B2 structure the magnon states are divided in two distinct branches, "acoustic" and "optical" with small temperature dependence of the peak positions. The width of the peak is inversely related to the magnon lifetime. From inspection, the width for the B2 structure at $T =$ 300 K is slightly larger than at $T =$ 10 K and thus giving shorter magnon lifetimes. However, the magnon lifetimes will also have a wave-vector dependence and a more involved analysis is needed. In the random alloy, the "acoustic" magnon branch is roughly located at the same energies as in the B2 structure, while the "optical" branch is significantly broadened in comparison. 

The most elaborate way to determine magnon lifetimes theoretically is through time dependent density functional theory and linear response\cite{Buczek2011}. Due to the complexity of such calculations, it has so far only been applied to elemental systems and not in alloys. Here, we therefore use an alternative simplified method to obtain the wave vector dependent magnon lifetimes $\tau(\mathbf{q})$. By fitting the dynamical structure factor for each wave vector with a Lorentzian and determine the full width half maximum (FWHM), $\Delta(\mathbf{q})$, the corresponding magnon lifetime is obtained through the relation $\tau(\mathbf{q})=\frac{2\hbar}{\Delta(\mathbf{q})}$. It is worth stressing that this approach only takes into account decay through Gilbert damping mechanism and not via Landau damping corresponding to electron-hole pair excitations within the Stoner continuum. However, for bulk materials as in this study it should be a reasonable good approximation, at least for the "acoustic" magnon branch.

\begin{figure}[hbp]
\includegraphics[width=9cm]{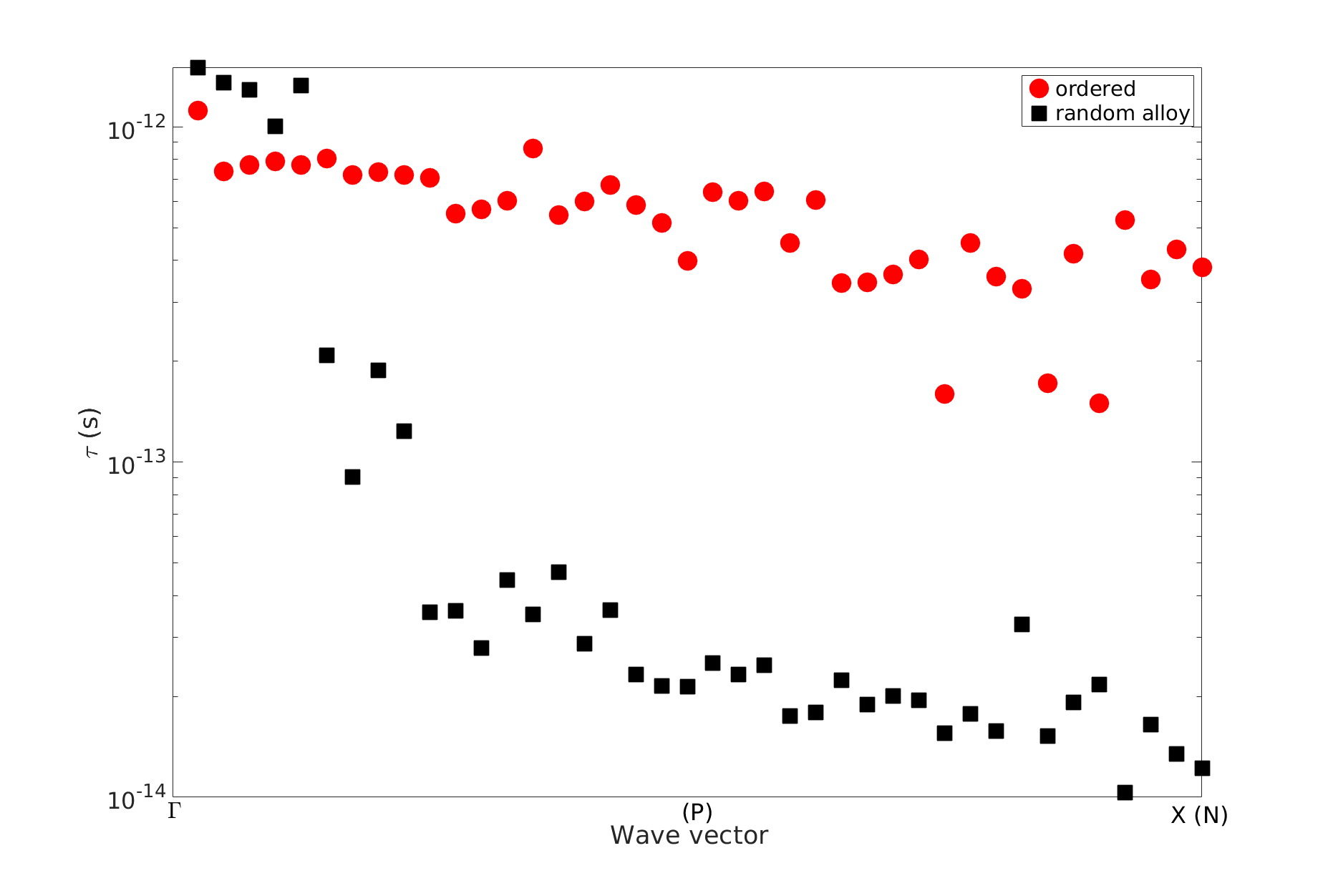}
\caption{Comparison of magnon lifetimes at $T =$ 300 K of FeCo in ordered B2 structure ($\Gamma-X$ direction) and as random alloy ($\Gamma-P-N$ direction).}
\label{fig:LTMag}
\end{figure}

In Fig.~\ref{fig:LTMag}, the calculated magnon lifetimes in specific directions of the Brillouin zone are displayed for both the ordered B2 structure and as random alloy. Due to the sensitivity of the fitting, the calculated lifetimes for each wave vector have an associated error bar that is of the order of the variation between neighbouring wave vectors values. Nevertheless, it is clear that on average the ordered B2 structure has longer magnon lifetimes compared to the random alloy. For the specific directions in Fig.~\ref{fig:LTMag}, the average magnon lifetime in B2 is approximately three times larger than for the random alloy (0.6 ps vs 0.2 ps). This behaviour is in line with what is normally expected from disordered systems, i.e. that increased disorder increases the scattering rates which effectively gives shorter quasi-particle lifetimes. A direct comparison can be made with the broadening of the electron bands in the spectral functions shown in Fig.~\ref{fig:BSF}.   

\section{Summary} \label{sec:summary}
We have presented magnon and finite temperature properties of random alloys using a combination of electronic structure calculations and atomistic spin dynamics simulations. Disorder is seen to have a pronounced effect on the magnon properties causing additional scattering and damping of magnon modes. However, the degree of magnon scattering and damping depends sensitively on the chemical composition of the alloy and also on the relative concentration of the constituent atomic species, prompting for material specific studies. For example, the magnon spectrum of permalloy (Fe$_{20}$Ni$_{80}$) is much more affected by disorder causing diffuse spectra in most of the Brillouin zone than that of Fe$_{75}$Co$_{25}$ where well-defined magnon states exist everywhere. Similarly, we compared the magnon properties of Fe$_{50}$Co$_{50}$, both as ordered structure and as random alloy. We found a distinct difference in the magnon density of states between the two as well as observing shorter magnon lifetimes in the random alloy. 
We hope that the present study will motivate new experiments in the next generation of neutron scattering facilities. These new facilities currently under construction may now have the required accuracy to fully resolve the intricate magnon features in random alloys.

\begin{acknowledgments}

Financial support from the Swedish Research Council (VR)(grant numbers VR 2015-04608, VR 2016-05980 and VR 2017-03763), the Swedish strategic research programme eSSENCE, and Swedish Energy Agency (grant number STEM P40147-1) is acknowledged. 

The computations were performed on resources provided by the Swedish National Infrastructure for Computing (SNIC) at the National Supercomputer Center (NSC), Link\"oping University, the PDC Centre for High Performance Computing (PDC-HPC), KTH, and the High Performance Computing Center North  (HPC2N), Ume\aa \ University.

\end{acknowledgments}

%

\end{document}